\documentclass[twocolumn]{aastex631}

\usepackage{natbib}
\usepackage{newtxtext,newtxmath}
\usepackage{amsmath}
\usepackage{graphics,graphicx}

\usepackage{xcolor}

\usepackage{mathtools}
\usepackage{enumitem}
\usepackage{lipsum}
\setlist[itemize]{leftmargin=*}

\usepackage{floatrow}
\usepackage[outercaption]{sidecap}    
\sidecaptionvpos{figure}{t}

\usepackage[T1]{fontenc}
\usepackage{ae,aecompl}

\shorttitle{Inferring dust and molecular gas masses from the dust continuum emission of quiescent high-z galaxies}
\shortauthors{R. K. Cochrane et al.}

\begin{document}
\title{Dust temperature uncertainties hamper the inference of dust and molecular gas masses from the dust continuum emission of quiescent high-redshift galaxies}

\author[0000-0001-8855-6107]{R. K. Cochrane}
\affiliation{Center for Computational Astrophysics, Flatiron Institute, 162 Fifth Avenue, New York, NY 10010, USA\\}

\author[0000-0003-4073-3236]{C. C. Hayward}
\affiliation{Center for Computational Astrophysics, Flatiron Institute, 162 Fifth Avenue, New York, NY 10010, USA\\}

\author[0000-0001-5769-4945]{D. Angl\'es-Alc\'azar}
\affiliation{Department of Physics, University of Connecticut, 196 Auditorium Road, U-3046, Storrs, CT 06269-3046, USA\\}
\affiliation{Center for Computational Astrophysics, Flatiron Institute, 162 Fifth Avenue, New York, NY 10010, USA\\}

\begin{abstract}
Single flux density measurements at observed-frame sub-millimeter and millimeter wavelengths are commonly used to probe dust and gas masses in galaxies. In this Letter, we explore the robustness of this method to infer dust mass, focusing on quiescent galaxies, using a series of controlled experiments on four massive haloes from the Feedback in Realistic Environments (FIRE) project. Our starting point is four star-forming, central galaxies at seven redshifts between $z=1.5$ and $z=4.5$. We generate modified quiescent galaxies that have been quenched for $100\,\rm{Myr}$, $500\,\rm{Myr}$, or $1\,\rm{Gyr}$ prior to each of the studied redshifts by re-assigning stellar ages. We derive spectral energy distributions for each fiducial and modified galaxy using radiative transfer. We demonstrate that the dust mass inferred is highly dependent on the assumed dust temperature, $T_{\rm{dust}}$, which is often unconstrained observationally. Motivated by recent work on quiescent galaxies that assumed $T_{\rm{dust}}\sim25\,\rm{K}$, we show that the ratio between dust mass and $1.3\,\rm{mm}$ flux density can be higher than inferred by up to an order of magnitude, due to the considerably lower dust temperatures seen in non star-forming galaxies. This can lead to an underestimation of dust mass (and, when sub-mm flux density is used as a proxy for molecular gas content, gas mass). This underestimation is most severe at higher redshifts, where the observed-frame $1.3\,\rm{mm}$ flux density probes rest-frame wavelengths far from the Rayleigh-Jeans regime, and hence depends super-linearly on dust temperature. We fit relations between ratios of rest-frame far-infrared flux densities and mass-weighted dust temperature that can be used to constrain dust temperatures from observations and hence derive more reliable dust and molecular gas masses. 
\end{abstract}
\keywords{Galaxy evolution (594) -- Galaxy quenching (2040) -- Radiative transfer simulations (1967) -- Submillimeter astronomy (1647)}
\section{Introduction}
The gas in galaxies is shaped by complex and connected cycles of physical processes, from the large-scale accretion of cool gas from the cosmic web onto halos, to black hole feeding and the collapse of molecular gas to young stars on small scales, as well as the feedback that ensues from these. Constraining the cosmic gas budget, as a function of gas phase, redshift and galaxy type, is therefore a key part of understanding the fuelling and quenching of star formation in galaxies over time (see \citealt{Peroux2020}, for a review). \\
\indent Since molecular hydrogen ($\rm{H}_{2}$), the direct fuel for star formation, does not have a permanent dipole moment, the mass of molecular gas must be inferred indirectly. In extragalactic studies, this is usually done using measurements of the rotational transitions of carbon monoxide ($^{12}\rm{CO}$, hereafter $\rm{CO}$) or the dust continuum emission. Observations of global $\rm{CO}(1-0)$ line fluxes at low redshift enabled by single-dish facilities such as Arecibo and the IRAM-30m telescope have enabled constraints on the gas-to-stellar mass fraction (hereafter, simply `gas fraction') as a function of stellar mass, as well as across the SFR-stellar mass plane \citep[e.g.][]{Saintonge2011a,Saintonge2016,Saintonge2017}. At higher redshifts, long integration times are required to measure $\rm{CO}(1-0)$ for a single source, and hence observed samples tend to comprise massive, sub-millimeter bright and/or gravitationally lensed sources \citep[e.g.][]{Danielson2011,Sharon2013,Sharon2016,Thomson2015}. At all redshifts, robust gas masses hinge on a reliable $M_{\rm{H}_{2}}$-to-$L_{\rm{CO}}$ conversion factor, $\alpha_{\rm{CO}}$, which is likely not universal but rather dependent on galaxy properties such as metallicity, gas temperature and velocity dispersion (see \citealt{Chiang2021} and \citealt{Bolatto2013}, for a review). At high redshifts, brighter, higher-order $\rm{CO}$ transitions are more accessible; these require additional assumptions about the relative luminosities of CO lines (see the review by \citealt{Carilli}). Despite these significant challenges, very careful analysis has enabled constraints on the evolution of molecular gas to $z\sim4$ \citep{Decarli2016,Decarli2019,Decarli2020}. These constraints appear to be in qualitative agreement with the evolution of the cosmic star formation history, displaying a decrease of a factor of $\sim6$ from $z\sim1.5$ to the present day. \\
\indent Use of the dust continuum emission to infer gas mass also has limitations. One method involves sampling the far-infrared (FIR) spectral energy distribution (SED) using multiple bands to derive the dust mass \citep[e.g.][]{Magdis2012,Dudzeviciute2019}. Inferring the cold gas mass is then subject to uncertainties on the gas-to-dust ratio, which is likely metallicity and redshift-dependent (see \citealt{Remy-Ruyer2014,DeVis2019} and references therein). An alternative is to extrapolate from a single measurement along the Rayleigh-Jeans (R-J) tail \citep{Scoville2014}. Since the R-J tail is nearly always optically thin, with a linear dependence of flux density on dust temperature, the dust mass (and then the gas mass, again assuming some uncertain conversion; \citealt{Liang2018,Privon2018}) may be derived given a dust emissivity per unit mass. This latter method is also dependent on the assumption of a single mass-weighted dust temperature, which is, on average, higher at high redshift \citep{Schreiber2018,Sommovigo2022}. Empirical scalings between CO-based gas mass and dust luminosity have been derived at low redshift (e.g. \citealt{Groves2015} found a linear scaling for the spiral galaxies in the KINGFISH survey; \citealt{Kennicutt2011}). However, it is not clear that these can be applied at higher redshift (e.g. \citealt{Decarli2016} applied the \citealt{Groves2015} scaling to galaxies at $z\sim1-2.5$ and obtained gas masses $\sim1.5$ times lower than CO-derived estimates).\\
\indent Most studies of ISM tracers have been limited to star-forming galaxies. Sub-millimeter observations of small numbers of individual high-z passive candidates have been performed, generally to confirm that they are indeed quiescent and not highly dust-reddened star-forming sources \citep[e.g.][]{Kalita2021,Santini2019,Santini2021}. Since quiescent galaxies tend to be faint at sub-mm wavelengths, population studies are limited to stacking analyses (e.g. \citealt{Gobat2018,Magdis2021}), which find considerably larger ISM masses than are observed in quiescent galaxies at $z\sim0$. \cite{Gobat2018} studied the stacked FIR emission for $977$ colour-selected quiescent galaxies with $\log_{10}(M_{\star}/\rm{M}_{\odot})>10.8$ at $1.4<z<2.5$ and fitted templates to the stacked dust SED. Their derived dust temperature is $21-25\,\rm{K}$, and their derived dust mass is $1^{+0.6}_{-0.4}\times10^{8}\,\rm{M}_{\odot}$. Assuming a gas-to-dust ratio (GDR) of 95, they inferred gas fractions of $5-10$ per cent. \cite{Magdis2021} performed a similar stacking analysis in redshift bins between $z=2.2$ and $z=0.3$. They constrained the luminosity-weighted dust temperatures of stacked samples of quiescent galaxies to be $<T_{\rm{dust}}>=21\pm2\,\rm{K}$, finding little evolution between $z=2$ and $z=0$. Assuming a constant solar-metallicity GDR of 92, they derived typical gas fractions for their stacked samples that drop from $\sim7-8$ per cent at $z=2$ to $\sim2$ per cent at $z=0.5$. There are, however, a number of uncertainties, including the dust-to-gas ratio. Local early type galaxies (ETGs) may have substantially lower dust-to-gas ratios than local late type galaxies, so assuming GDRs of $\sim100$ may not be valid \citep{Kokusho2019}.\\
\indent Stacked samples can also suffer contamination. Photometrically selected samples of quiescent galaxies can be contaminated by dusty star-forming galaxies, either due to mis-classification of small numbers of sources or from neighbouring galaxies via source blending in the lower spatial resolution FIR data. Although observationally expensive, studies of individual galaxies mitigate these risks. Small numbers of existing deep single-source observations present a diverse picture of the gas content of high-redshift quiescent galaxies. \cite{Williams2021} present CO(2-1) observations of six massive ($\log_{10}(M_{\star}/\rm{M}_{\odot})>11.3$) quiescent galaxies at $z\sim1.5$, five of which were undetected, and were able to place ($3\sigma$) upper limits of gas fractions at $2-6$ per cent. These gas fractions are lower than inferred from the stacking analysis of dust continuum emission presented by \cite{Gobat2018}. They are also lower than studies of post-starburst galaxies, which appear to harbour substantial molecular gas reservoirs (though little dense gas; \citealt{French2018}), despite their low star formation rates \citep{French2015,Suess2017,Bezanson2021}. \\
\indent Expanding samples of individual quiescent galaxies targeted with ALMA, \cite{Whitaker2021a} presented $1.3\,\rm{mm}$ observations of six strongly-lensed sources from the REsolving QUIEscent Magnified (REQUIEM) galaxy survey \citep{Akhshik2020}. The galaxies have redshifts in the range $z=1.6-3.2$, and all but one has a very low star formation rate, as derived from ultraviolet-to-near-infrared SED fitting (the highest redshift source has an SFR in line with the main sequence at this redshift. However, \cite{Man2021} find that the SFR derived from SED fitting is higher than that derived from [O{\small{II}}], and argue that this is indicative of recent, rapid quenching). Four of the sources were undetected in the ALMA observations, down to rms values of $9-56\,\mu\rm{Jy}$. \cite{Whitaker2021a} follow the methods developed by \cite{Scoville2014}, converting limits on $S_{1.3}\,\rm{mm}$ to limits on dust mass, assuming a dust temperature of $25\,\rm{K}$. Converting this to an upper limit on gas mass using a GDR of $100$, they derive upper limits on gas fractions of $\sim1$ per cent for the four undetected sources and argue that these very low gas masses imply rapid, early depletion of gas. In a subsequent paper, \cite{Whitaker2021b} caution that their choice of GDR may be inappropriate for the types of galaxies within their sample; in the cosmological hydrodynamical simulation SIMBA \citep{Dave2019}, the GDRs of quiescent galaxies span six orders of magnitude.\\
\indent In this Letter, we investigate a secondary, unstudied aspect: the robustness of inferring the dust mass from $S_{1.3\,\rm{mm}}$ for quiescent, high-redshift sources. Two linked aspects of this may weaken the robustness of $S_{1.3\,\rm{mm}}$ as a dust mass tracer. The first is the impact of a reduction in recent star formation (and hence less dust heating) on the FIR emission of a galaxy. The second potential issue is the fixed observed-frame wavelength; at high redshifts, we probe a shorter rest-frame wavelength and hence move further from the R-J regime, where the flux density depends linearly on temperature. In Section \ref{sec:FIRE_sims}, we introduce the hydrodynamical simulations and radiative transfer methods used to create mock SEDs for star-forming and quenched galaxies. In Section \ref{sec:test_inference} we infer dust masses for the modelled sources from their $1.3\,\rm{mm}$ flux densities, assuming a dust temperature of $25\,\rm{K}$. In combination with the `ground truth' known from the simulations, we characterise the reliability of these estimates as a function of redshift. In Section \ref{sec:two_submm_measurements}, we show how estimates of more realistic dust temperature can be made to improve constraints on the dust mass. We draw our conclusions in Section \ref{sec:conclusions}.

\begin{figure*} 
\centering
\vspace{0.2cm}
\includegraphics[scale=0.47]{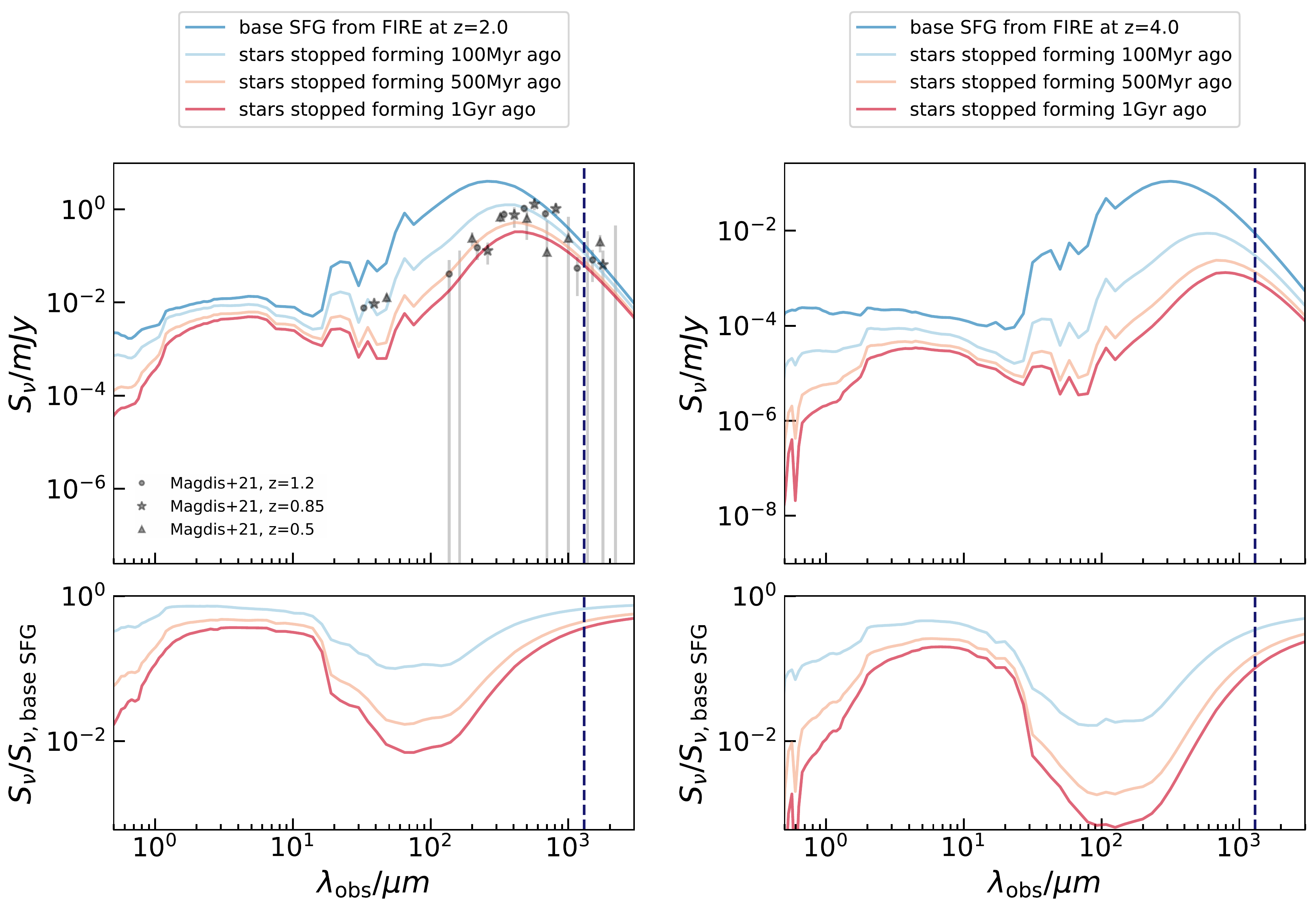}
\vspace{0.2cm}
\caption{The impact of quenching star formation on a galaxy's observed-frame SED. Upper panels: dark blue lines show the SED generated for a simulated star-forming galaxy at $z=2.0$ (left) and $z=4.0$ (right). Other lines show the impact of a controlled `quenching' event: the pale blue lines show the SED generated if the galaxy had no stars formed within $100\,\rm{Myr}$ of $z=2.0$ (or $z=4.0$); the pale and dark red lines show the same for no stars formed within the prior $500\,\rm{Myr}$ and $1\,\rm{Gyr}$, respectively. {We overlay (appropriately redshifted) data from stacked samples of observed colour-selected massive quiescent galaxies at $z\sim0.3-1.4$ \citep{Magdis2021} on the panel that shows the SED for the lower-redshift snapshot. The data are in qualitative agreement with our predicted SEDs.} Lower panels show the ratio of emission at different wavelengths, relative to the fiducial case of the star-forming source. Dashed navy lines mark an observed-frame wavelength of $1.3\,\rm{mm}$. For the quenched galaxies, the radiation field is softer due to the absence of young, massive stars and the dust temperature is lower. At $z=2.0$, $1.3\,\rm{mm}$ in the observed frame is near the R-J regime, where the flux depends linearly on dust temperature. Hence, the difference in dust temperature between star-forming and quenched sources does not result in a substantial reduction in observed $1.3\,\rm{mm}$ flux density. At $z=4$, $1.3\,\rm{mm}$ probes a shorter rest-frame wavelength ($260\,\mu\rm{m}$), and modest temperature differences are amplified because of the super-linear dependence of flux density on dust temperature.\\}
\label{fig:seds_A2}
\end{figure*}

\section{The FIRE simulations}\label{sec:FIRE_sims}
\subsection{Base simulations of massive, star-forming galaxies}
Our base simulations are presented in detail in \cite{Cochrane2022}, and we provide only a brief summary here. The FIRE project \citep{Hopkins2014,Hopkins2017} is a set of state-of-the-art hydrodynamical cosmological zoom-in simulations that explore the role of stellar feedback in galaxy formation and evolution. In this Letter, we perform controlled experiments on central galaxies of four haloes presented by \cite{Angles-Alcazar2017}, who implemented the FIRE-2 physics model \citep{Hopkins2017} and a novel on-the-fly treatment for the seeding and growth of supermassive black holes (SMBHs) developed by \cite{Angl2018}. These haloes had originally been simulated by \citet{Feldmann2016, Feldmann2017a} with the original FIRE model \citep{Hopkins2014} as part of the {\sc MassiveFIRE} suite. The four haloes were selected to have dark matter halo masses of $M_{\rm{halo}}\sim10^{12.5} \,\rm{M}_{\odot}$ at $z=2$. The central galaxies of these haloes have stellar masses of $\sim7\times10^{10}-3\times10^{11}\,\rm{M}_{\odot}$ and star formation rates of $\sim50-200\,\rm{M}_{\odot}\rm{yr}^{-1}$ at $z=2$, with a variety of formation histories; see \cite{Feldmann2017a} for details. Note that these simulations do not include feedback from the accreting SMBHs. \\
\indent For the purposes of illustrating redshift trends in this Letter we focus on a subset of snapshots at $z=1.5$, $z=2.0$, $z=2.5$, $z=3.0$, $z=3.5$, $z=4$ and $z=4.5$. 

\subsection{Controlled quenching experiments}\label{sec:modifications}
We perform controlled experiments on the four central galaxies to investigate the impact of quenching star formation on the observable dust continuum emission. For each galaxy at each of the seven redshifts studied, we mimic the quenching of star formation $100\,\rm{Myr}$, $500\,\rm{Myr}$ or $1\,\rm{Gyr}$ prior. In practice, this means identifying the stars that formed within the relevant epoch (within $\Delta t_{\rm q} = [0.1,0.5,1]\,\rm{Gyr}$), and replacing their ages with the corresponding value of $\Delta t_{\rm q}$. This is equivalent to assuming that recent stars formed within a single burst. We then predict multi-wavelength emission again, following the same methods described in Section \ref{sec:synthetic_obs}. In each case, we maintain the same galaxy gas and dust content and the same total stellar mass. The only difference between fiducial and modified galaxies is the presence (or absence) of the radiation field of the young stellar populations. We choose not to model these differences within this study; our `controlled experiment' setup is optimised to test the impact of the quenching of recent star formation on the SED most cleanly, {without dust mass as an additional varying parameter.\\}
{In reality, we expect there to be differences between the gas and dust content of real quiescent and star-forming galaxies. Indeed, the positions of low-redshift galaxies on the stellar mass-star formation rate plane can mostly be explained by their cold gas reservoirs \citep[][]{Saintonge2016}, with specific star formation rate showing a tight correlation with molecular gas fraction \citep{Spilker2018}. However, as discussed in the Introduction, the distribution of gas and dust fractions of quiescent galaxies at high redshift are more poorly constrained. Because of this, and because the aim of this Letter is to test a method rather than put constraints on dust fractions, we do not vary dust mass along with the star-formation history. We confirmed that our results are robust to variations in assumed dust fraction by repeating the experiment with a significantly decreased dust mass (10\% of the fiducial value). We discuss this in Section \ref{sec:test_inference}.}

\subsection{Synthetic observations with {\small{SKIRT}}}\label{sec:synthetic_obs}
The primary aim of this Letter is to illustrate the differences in millimeter flux-to-dust mass ratio between high-redshift galaxies that are star-forming and those that are quenched. We first model the observed-frame SEDs for both the fiducial FIRE simulations of star-forming galaxies and our modified quenched galaxies. Following \cite{Cochrane2019}, \cite{Parsotan2021} and \cite{Cochrane2022}, we use the {\small{SKIRT}} radiative transfer code \citep{Baes2011,Camps2014} to make predictions for the rest-frame ultra-violet to far-infrared SEDs of each galaxy snapshot. For both fiducial and modified snapshots, gas and star particles within $0.1R_{\rm{vir}}$ are drawn directly from the FIRE-2 simulation. The only difference is that, for the modified snapshots, stars formed after the set quenching time are modified within the star particle data file, assuming that they formed in a single burst $100\,\rm{Myr}$, $500\,\rm{Myr}$ or $1\,\rm{Gyr}$ prior, and star formation was halted thereafter. The total stellar mass of a modified snapshot is the same as that of its parent star-forming galaxy. We do not alter the gas content of the galaxy or halo, so the gas mass and metallicity of fiducial and modified galaxies are the same, for a given halo and redshift. Dust particles are assumed to follow the distribution of the gas particles, with a dust-to-metals mass ratio of $0.4$ \citep{Dwek1998,James2002}. We assume dust destruction at $>10^6\,{\rm K}$ \citep{Draine1979,Tielens1994}. We model a mixture of graphite, silicate and PAH grains using the \cite{Weingartner2001} Milky Way dust prescription. Star particles are assigned \citet{Charlot2003} SEDs based on their ages and metallicities. We perform the radiative transfer on an octree dust grid, in which cell sizes are adjusted according to the dust density distribution, with the condition that no dust cell may contain more than $0.0001\%$ of the total dust mass of the galaxy.

\begin{figure} 
\centering
\vspace{0.2cm}
\includegraphics[scale=0.7]{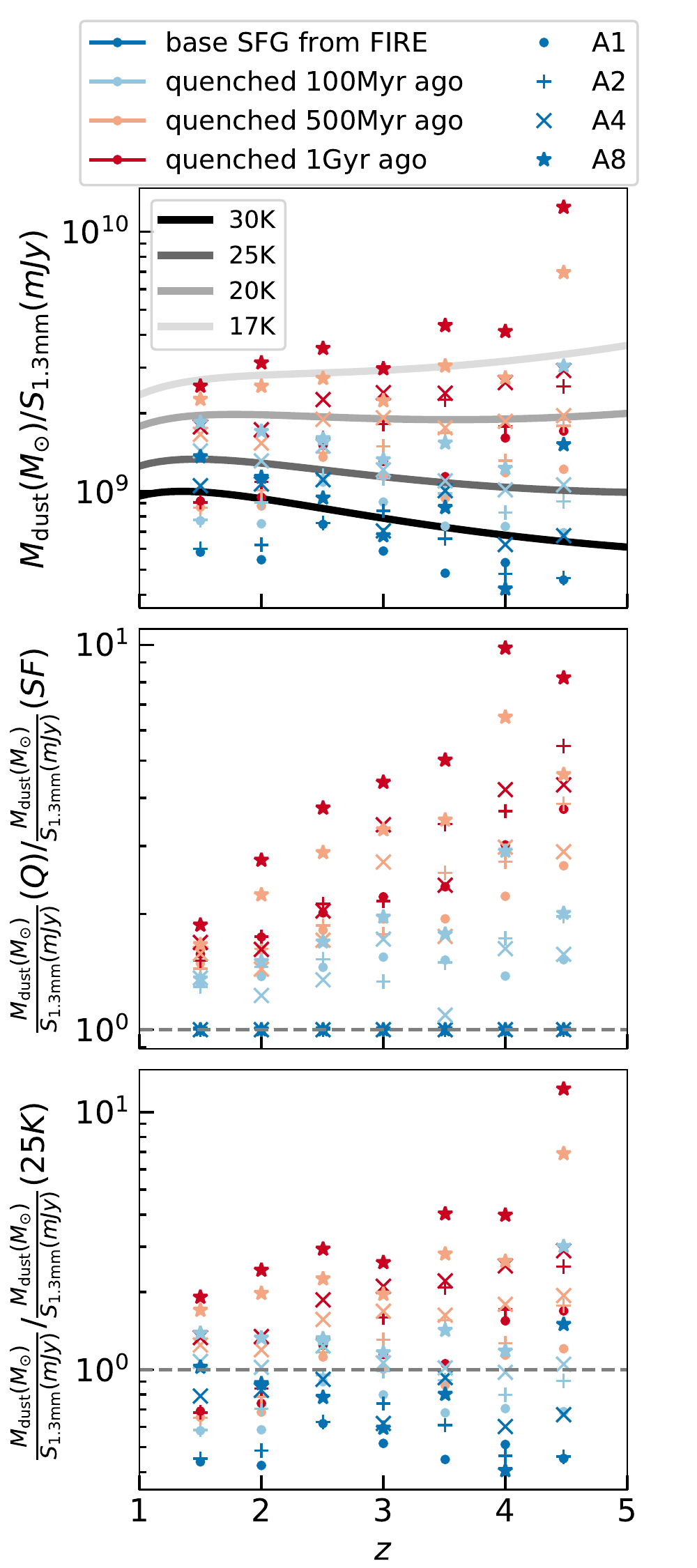}
\vspace{0.2cm}
\caption{Upper: the $M_{\rm{dust}}$-to-$S_{\rm{1.3mm}}$ ratio (MSR) derived from our radiative transfer calculations, with predictions for the ratio at different dust temperatures overlaid (see Equation \ref{eq:fiducial_case}). Middle: the predicted MSR for different quenching scenarios, divided by that calculated for the fiducial star-forming source, for each galaxy at a given redshift. Lower: the calculated MSR divided by the canonical ratio (Equation \ref{eq:fiducial_case}), assuming a dust temperature of $25\,\rm{K}$. For our fiducial sample of star-forming galaxies (dark blue), the true dust mass per unit flux density tends to be slightly lower than predicted, since the mass-weighted dust temperature is actually higher than $25\,\rm{K}$. For the quenched galaxies, the true scaling between dust mass and $1.3\rm{mm}$ flux density is higher than the canonical ratio by factors of a few. This effect is more pronounced at higher redshifts and can lead to underestimation of dust (and gas) masses from continuum flux density measurements for galaxies with little or no recent star formation by as much as an order of magnitude.\\}
\label{fig:underprediction}
\end{figure}

\subsection{The impact of controlled quenching on a galaxy's SED}\label{sec:sed_comparison}
In Figure \ref{fig:seds_A2} we present examples of the observed-frame SEDs generated by the radiative transfer modelling for one FIRE galaxy (the central galaxy of halo A8) at two redshifts, $z=2.0$ (left) and $z=4.0$ (right). In addition to the SED of the fiducial star-forming galaxy (shown in dark blue), we show the SED predicted for the controlled quenching scenarios (representing, from light blue to dark red, no stars formed within the last $100\,\rm{Myr}$, $500\,\rm{Myr}$ and $1\,\rm{Gyr}$). We also show the ratio between the emission predicted for the controlled quenching scenarios and the fiducial star-forming case (see the lower panels). While we have predicted the observed-frame optical to millimeter emission of the galaxies, in this Letter we focus on the measurable $1.3\,\rm{mm}$ emission (marked on the SEDs with a dashed vertical line). We choose this wavelength following the recent work of \cite{Whitaker2021a}. \\
\indent For the galaxy snapshot at $z=1.5$, the effects of quenching star formation on the $1.3\,\rm{mm}$ emission are modest. There is a small shift in the peak of the dust SED to longer wavelengths, reflecting the colder temperature of the dust due to the softer radiation field in the absence of young, massive stars. However, since the observed $1.3\,\rm{mm}$ emission probes near the R-J tail, the flux density depends approximately linearly on temperature, and so the modest difference in dust temperature does not result in a substantial difference in flux density. At higher redshifts, the observed-frame $1.3\,\rm{mm}$ flux density probes further from the RJ tail, nearer to the peak of the dust SED; here, modest differences in dust temperature are amplified due to the super-linear dependence of the flux density on temperature. For example, for the $z=4$ galaxy shown in Figure \ref{fig:seds_A2}, $S_{1.3\,\rm{mm}}$ drops by up to a factor of $\sim5$, despite the dust and gas mass remaining constant by construction. As a result, the ratio of flux density to dust (or gas) mass will vary strongly with dust T, which is sensitive to the recent star formation history. We explore this in Section \ref{sec:test_inference}.

\section{Inferring dust mass from sub-mm flux density}\label{sec:Mdust_inference}
\subsection{The robustness of the dust mass inferred from $S_{1.3\,\rm{mm}}$}\label{sec:test_inference}
The dust continuum emission of a galaxy is frequently parameterised as follows, for the case where the dust is optically thin at the rest-frame frequency (e.g. \citealt[][]{James2002,Hayward2011,Scoville2017}):
\begin{equation}\label{eq:Mdust_eq}
S_{\rm{\nu_{\rm{obs}}}} = (1+z) \kappa_{\nu_{\rm{rest}}}B_{\nu_{\rm{rest}}}\left(T_{\rm{dust}}^{\rm{BB}}\right) \frac{M_{\rm{dust}}}{D_{L}^{2}}
\end{equation}
where $S_{\rm{\nu_{\rm{obs}}}}$ is the observed flux density of the source, $z$ is its redshift, $D_{L}$ is the corresponding luminosity distance,  $\nu_{\rm{rest}}$ is the rest-frame frequency, $B_{\nu_{\rm{rest}}}$ is the Planck function, $T_{\rm{dust}}^{\rm{BB}}$ is the effective dust temperature, and $M_{\rm{dust}}$ is the total dust mass. $\kappa_{\nu_{\rm{rest}}}$ is the dust opacity per unit dust mass (also known as the mass absorption coefficient). This follows a power law in frequency as: $\kappa_{\nu_{\rm{rest}}} = \kappa_{\nu_{\rm{ref}}} \left(\frac{\nu_{\rm{rest}}}{\nu_{\rm{ref}}}\right)^{\beta}$, where $\kappa_{\rm{ref}}$ is a measured value at some reference frequency, $\nu_{\rm{ref}}$. We use $\kappa_{\nu_{850\,\mu\rm{m}}}=0.0484\,\rm{m}^{2}\rm{kg}^{-1}$ as our reference. We assume a dust emissivity index $\beta=1.8$, for consistency with \cite{Whitaker2021a}. \\
\indent A simple re-arrangement gives the ratio of dust mass to observed flux density:
\begin{equation}\label{eq:fiducial_case}
\frac{M_{\rm{dust}}}{S_{\rm{\nu_{\rm{obs}}}}} = \frac{1}{1+z} \frac{D_{L}^{2}}{\kappa_{\nu_{\rm{rest}}}B_{\nu_{\rm{rest}}}\left(T_{\rm{dust}}^{\rm{BB}}\right)}.
\end{equation}\\
Dust mass can thus be inferred from the observed flux density, and the molecular gas mass is often extrapolated assuming some constant gas-to-dust mass ratio (typically 100, for solar metallicity gas).\\
\indent In the R-J regime, where $\lambda_{\rm{rest}}>>\frac{hc}{k_{B}T}$, $B_{\nu_{\rm{rest}}}\left(T_{\rm{dust}}^{\rm{BB}}\right)=\frac{2\nu_{\rm{rest}}^{2}{k_{B}T}}{c^{2}}$, and hence:
\begin{equation}\label{eq:R-J_case}
\frac{M_{\rm{dust}}}{S_{\rm{\nu_{\rm{obs}}}}} = \frac{D_{L}^{2}}{2(1+z)} \frac{c^{2}}{\nu_{\rm{rest}}^{2}{k_{B}T}\kappa_{\nu_{\rm{rest}}}}.
\end{equation}\\
Hence, when in both the optically-thin and R-J regimes, $M_{\rm{dust}}/S_{\nu}\propto 1/T$.\\
\indent In this Letter, we explore how the dust mass to flux density ratio, calculated at an observed wavelength of $1.3\,\rm{mm}$ (hereafter the `MSR'), changes with redshift and recent star formation history. In the upper panel of Figure \ref{fig:underprediction}, we show the MSR for each simulated galaxy, at seven redshifts in the range $z=1.5$ to $z=4.5$. Galaxies that have been quenched for longer have lower MSRs, as expected since the absence of young, massive stars causes a softer radiation field and a lower dust temperature. We overplot the expected MSR, calculated using Equation \ref{eq:fiducial_case}, for four effective dust temperatures. The fiducial star-forming galaxies tend to have MSRs consistent with $T_{\rm{dust}}\sim30\,\rm{K}$, while galaxies quenched $500\,\rm{Myr}-1\,\rm{Gyr}$ prior to the snapshot have MSRs consistent with effective temperatures as low as $\sim17-20\,\rm{K}$. In the middle panel of Figure \ref{fig:underprediction}, we quantify the ratio between the MSR of each quenched galaxy and the MSR of the fiducial galaxy at each redshift. This shows the boosting of the MSR relative to the star-forming case. This can be as large as a factor of $10$. In the lower panel of Figure \ref{fig:underprediction}, we compare the MSR to that expected from Equation \ref{eq:fiducial_case}, assuming a dust temperature of $25\,\rm{K}$ in line with \cite{Scoville2014,Scoville2017} and the recent work of \cite{Whitaker2021a}.\\
\indent At $z=1.5$ (just lower than the lowest redshift of the \citealt{Whitaker2021a} sample), the MSR is equal to the expected value to within a factor of $\sim2$. This is expected intuitively from Figure \ref{fig:seds_A2}: at this redshift, observed-frame $1.3\,\rm{mm}$ emission probes rest-frame $520\,\mu\rm{m}$, near to the R-J tail, and is dominated by cool dust in thermal equilibrium. Small deviations from the ratio predicted by Equation \ref{eq:fiducial_case} will result from different dust temperatures, as $S_{\nu}/M_{\rm{dust}}\propto T$ in the R-J regime (see Equation \ref{eq:R-J_case}). As redshift increases, discrepancies between the true MSR and that calculated using Equation \ref{eq:fiducial_case} become more marked.  As shown in the middle panel of Figure \ref{fig:underprediction}, the MSR becomes more dependent on temperature with redshift, exceeding that derived  assuming $25\,\rm{K}$ by factors of up to $10$ at $z>4$. This is because we have used a fixed observed-frame wavelength, and for our high-redshift quenched sources, we are not probing the R-J tail. For the R-J regime, we require $\lambda_{\rm{rest}}>>\frac{hc}{k_{B}T}$, i.e. $\lambda_{\rm{rest}}/\rm{mm}>>14.4/T(\rm{K})$. For dust temperatures of $25\,\rm{K}$, this requires  $\lambda_{\rm{rest}}>>580\,\mu\rm{m}$, and for temperatures of $20\,\rm{K}$, this requires $\lambda_{\rm{rest}}>>720\,\mu\rm{m}$. At $z=1.5$, the $1.3\,\rm{mm}$ flux density probes rest-frame $520\,\mu\rm{m}$; for R-J, dust temperatures must be $>>27.7\,\rm{K}$. At higher redshifts and for lower dust temperatures, the rest-frame wavelength probed by $1.3\,\rm{mm}$ flux density is far from the R-J regime (see also the right-hand panel of Figure \ref{fig:seds_A2}), and the flux density depends super-linearly on dust temperature. We demonstrate this in Figure \ref{fig:underprediction}; for high-redshift quenched galaxies without a constrained dust temperature, $1.3\,\rm{mm}$ flux density places weak constraints on the dust (or gas) mass.\\
\indent {To test the robustness of our conclusions to significant changes in dust mass, we repeat the experiment with dust mass reduced to $10\%$ of the original value. This leads to an increase in the mass-weighted temperature of the dust by several $\rm{K}$ and a decrease  in the amplitude and peak wavelength of the dust SED. However, when we normalise by the dust mass, as we have in this Letter, the difference is very small. Our conclusions are thus insensitive to the exact values of dust mass in high redshift quiescent galaxies.}\\
\indent {Our results are consistent with the observationally-driven conclusions of \cite{Magdis2021}. They constrain the average luminosity-weighted temperature of colour-selected samples of massive ($M_{\star}\sim10^{11}\,\rm{M_{\odot}}$) quiescent galaxies to be $21\pm2\,\rm{K}$ out to $z\sim2$. They show that following the method presented by \cite{Scoville2017} and assuming a mass-weighted temperature of $25\,\rm{K}$ leads to an underestimation of gas mass compared to their template-based method. This underestimation factor is $\sim2-4$, in good agreement with the theoretical work presented here.}

\begin{figure*} 
\vspace{0.2cm}
\includegraphics[width=\columnwidth]{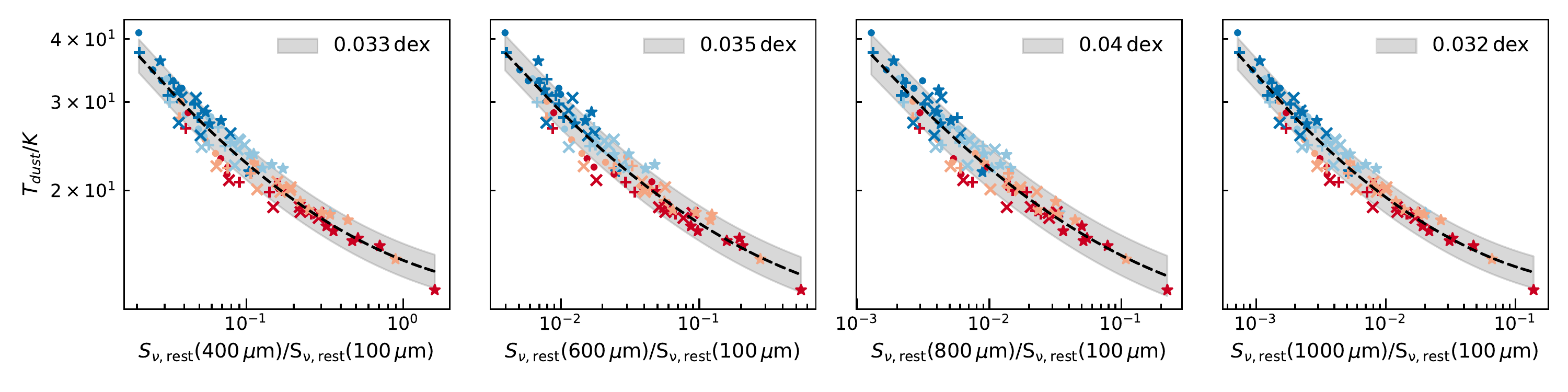}
\caption{Mass-weighted dust temperature versus the ratio of two rest-frame flux densities. Colours and symbols are the same as used and defined in Figure \ref{fig:underprediction}. The tight correlations show that measurements of two flux densities, or one measurement and an upper limit, can put constraints on the dust temperature, thereby enabling more accurate inference of the dust (and thus the gas) mass. We overplot the best-fitting relations derived for each wavelength combination (black dashed line; see Equation \ref{eq:T_relation}). The shaded region shows the $1\,\sigma$ scatter between the true mass-weighted temperature and the best-fitting relation.\\}
\label{fig:temperature_vs_ratio}
\end{figure*}

\begin{figure*}
\vspace{-0.2cm}
    \includegraphics[width=0.9\textwidth]{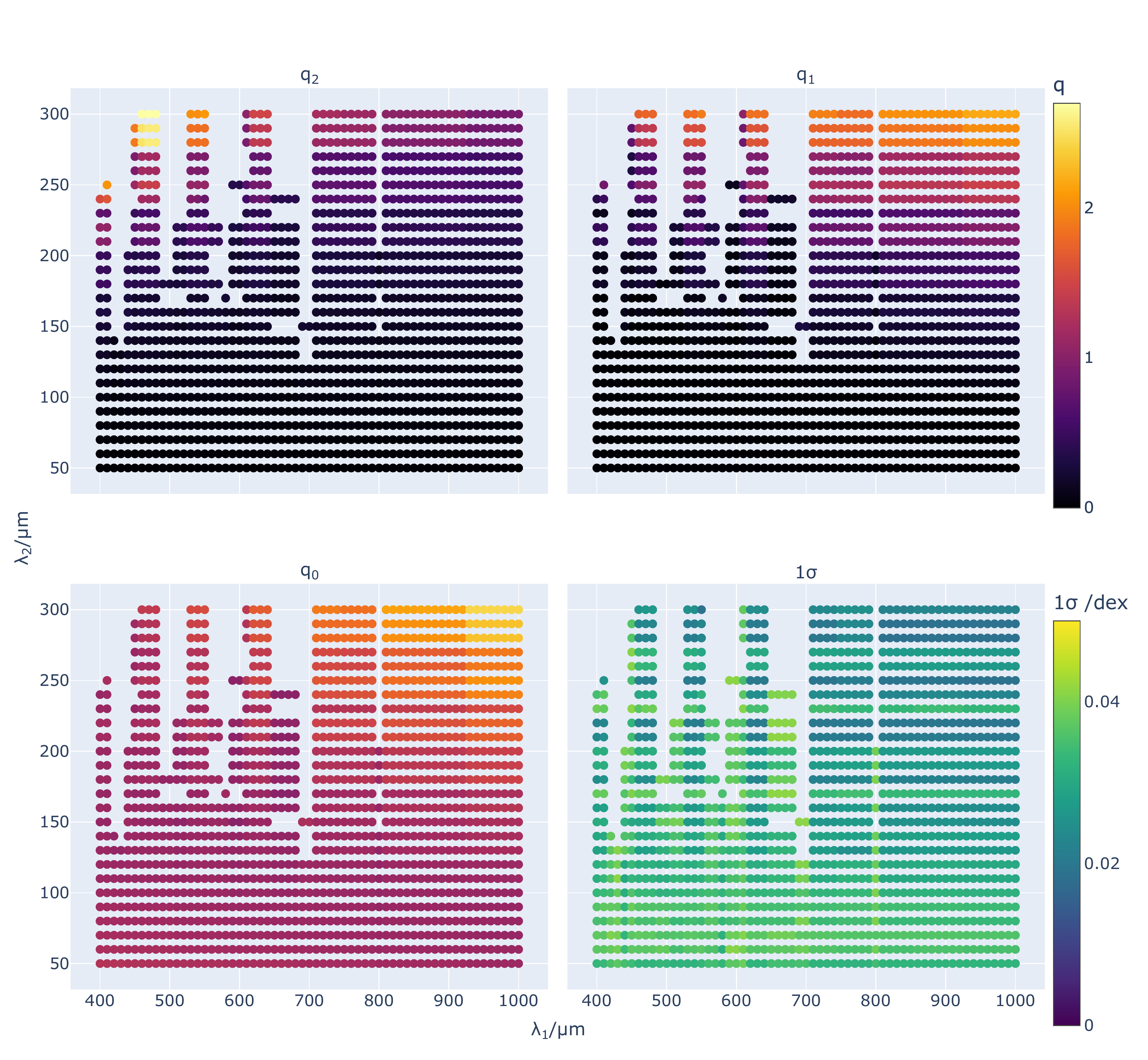}
\caption{Panels (a), (b) and (c): best-fitting coefficients $q_{2}$, $q_{1}$ and $q_{0}$ from Equation \ref{eq:T_relation}, fitted using different combinations of rest-frame wavelengths $\lambda_{1}$ and $\lambda_{2}$. Panel (d) shows the $1\,\sigma$ scatter in the fitted relation, calculated using all snapshots. The scatter characterises how robustly the mass-weighted temperature can be inferred from the two flux density measurements taken at $\lambda_{1}$ and $\lambda_{2}$. We only show combinations for which the fitted formula predicts the dust temperature for $68$ per cent of snapshots to within $10$ per cent ($0.041\,\rm{dex}$) of the true value. The lowest values of scatter are found for longer wavelengths (i.e. in the upper right-hand corner of the plots). The exact values of the coefficients (available from the online version of this Letter) should help constrain dust temperatures of observed galaxies when two sub-mm/mm flux densities are available.
A plot.ly-based interactive figure is available in the online Journal. The interactive version allows the reader to look up individual values of the coefficients by mousing over the individual scatter plot(s). The data and scripts used to make this interactive figure are available as data-behind-the-Figure.\\}
\label{fig:coefficients_Tfit}
\end{figure*}

\subsection{Constraining the dust temperature using two sub-mm flux density measurements}\label{sec:two_submm_measurements}
As shown in Section \ref{sec:test_inference}, the dependence of the sub-mm flux density on the dust temperature causes the MSR to vary depending on the recent SFH of the galaxy, especially at high redshift. In this section, we show that combining measurements at two rest-frame FIR wavelengths can put constraints on the dust temperature of the galaxy (and hence enable better constraints on the dust mass). \footnote{The effective temperatures inferred from the MSR and the mass-weighted dust temperatures derived directly from the simulations (calculated from the dust grid on which radiative transfer is performed) show excellent agreement. Hence, we focus on constraining the mass-weighted dust temperature here.} Far-infrared SEDs of observed galaxies are generally well-fitted by a modified black body. Assuming this form, the ratio of two rest-frame flux densities, $S_{\nu_{\rm{rest,1}}}$ and $S_{\nu_{\rm{rest,2}}}$ is expected to be: 
\begin{equation}\label{eq:S_ratios}
\frac{S_{\nu_{\rm{rest,1}}}}{S_{\nu_{\rm{rest,2}}}} = \frac{(1-\exp[-\tau_{\nu_{\rm{rest,1}}}])B_{\nu_{\rm{rest,1}}}(T_{\rm{dust}}^{\rm{BB}})}{(1-\exp[-\tau_{\nu_{\rm{rest,2}}}])B_{\nu_{\rm{rest,2}}}(T_{\rm{dust}}^{\rm{BB}})},
\end{equation}
where the optical depth $\tau_{\nu}$ is frequently parametrised as $\tau=(\nu/\nu_{0})^{\beta}$ and $\nu_{0}$ is a constant that relates to the wavelength at which the medium becomes optically thick. In the limit of long wavelengths, $\exp(-\tau_{\nu_{\rm{rest}}})\approx\tau_{\nu_{\rm{rest}}}$, $B_{\nu_{\rm{rest}}}(T)\approx\frac{2\nu_{\rm{rest}}^{2}k_{B}T}{c^{2}}$; the ratio of two long wavelength flux densities becomes dependent on the two frequencies and the dust emissivity index, $\beta$, alone. When this condition is not satisfied, the flux density ratio also depends on the dust temperature. Here, we consider combinations of rest-frame wavelengths that include one that is firmly outside the R-J regime, to put constraints on the dust temperature. {Note that our $S_{\nu_{\rm{rest,1}}}$ and $S_{\nu_{\rm{rest,2}}}$ do not include a correction to account for the CMB as an observing background. When inferring flux densities from observations, corrections to account for this should be made as usual (see the derivations provided by \citealt{DaCunha2013} and their application in \citealt{Magdis2021}, for example).} \\
\indent We show the correlation between mass-weighted dust temperature (calculated directly from the dust grid on which the radiative transfer is performed) and ratios of flux densities at two rest-frame FIR/sub-mm wavelengths (drawn from our modelled SEDs) in Figure \ref{fig:temperature_vs_ratio}. Deriving the expected relationship from Equation \ref{eq:S_ratios} requires assumptions about $\beta$, $\nu_{0}$ and the dust temperature distribution. While this can be performed as required for any choice of these parameters, we present here empirical fits to the data derived from our simulation. These implicitly fold in the dust temperature distribution at different mass-weighted dust temperatures, $T_{\rm{MW}}$ \footnote{We find that the distribution of dust temperatures within both star-forming and quiescent FIRE galaxies can be well-fitted by a normal distribution in $\log_{10}({T[\rm{K}}])$, i.e. $f(\log_{10}({T}))=\frac{1}{\sigma\sqrt{2\pi}}\exp{\Big[-\frac{1}{2}\Big(\frac{\log_{10}{T}-\log_{10}{T_{\rm{MW}}}}{\sigma}\Big)^{2}\Big]}$, with mean equal to the mass-weighted temperature, $T_{\rm{MW}}$, and $\sigma$ a function of this temperature. Galaxies with higher mass-weighted temperatures tend to have larger values of $\sigma$. Fitting $\sigma$ as a function of $T_{\rm{MW}}$ yields $\sigma = 0.00327\times \log_{10}(T_{\rm{MW}}[\rm{K}])+0.00271$.}. As shown by the dashed black line in each panel of Figure \ref{fig:temperature_vs_ratio}, our data are well-fitted by a relation of the form: 
\begin{equation}\label{eq:T_relation}
\log_{10}(T_{\rm{MW}}[\rm{K}]) = q_{2} \left(\log_{10}\left(\frac{S_{\nu_{\rm{rest,1}}}}{S_{\nu_{\rm{rest,2}}}}\right)\right)^{2}\,+\, q_{1}\log_{10}\left(\frac{S_{\nu_{\rm{rest,1}}}}{S_{\nu_{\rm{rest,2}}}}\right)\,+\,q_{0}
\end{equation}
We show the fitted coefficients $q_{2}$, $q_{1}$ and $q_{0}$ for different combinations of long and short wavelengths in panels (a), (b) and (c) of Figure \ref{fig:coefficients_Tfit} (see the online version of this Letter for exact numerical values). In panel (d), we show the $1\,\sigma$ scatter in the fitted relation. This was calculated using all snapshots, by comparing the temperature derived from two flux density data points and the relation to the actual mass-weighted dust temperature drawn directly from the simulations. In Figures (a), (b), and (c), we only show combinations of wavelengths for which this scatter is $<0.041\,\rm{dex}$; this corresponds to the dust temperature being predicted to within $10$ per cent of the true value for $68$ per cent of snapshots. Given that the relations between combinations of flux densities and mass-weighted dust temperature are in general so tight, we make the coefficients $q_{2}$, $q_{1}$ and $q_{0}$ available in the online version of this Letter. We propose that they might be used in observational studies to put constraints on dust temperatures of observed galaxies. 

\section{Conclusions}\label{sec:conclusions}
In this Letter, we have performed controlled quenching experiments on four simulated massive, star-forming galaxies from the FIRE zoom-in simulations at $z=1.5-4.5$. We study the impact of the cessation of star formation on the galaxies' SEDs, focusing particularly on the change in observed-frame $1.3\,\rm{mm}$ flux densities. Despite maintaining constant dust mass for the fiducial star-forming and modified quiescent galaxies, the $1.3\,\rm{mm}$ flux density decreases after quenching by up to an order of magnitude, primarily due to the decrease in dust temperature owing to the absence of young, massive stars, which results in a softer radiation field. This effect is particularly strong at high redshifts, where a shorter rest-frame wavelength is probed and $1.3\,\rm{mm}$ flux density probes far from the R-J regime and depends super-linearly on dust temperature. Our results have implications for observational studies of quenched galaxies, where dust mass is frequently inferred from a single millimeter-wavelength flux density. We highlight the importance of using a realistic dust temperature in such modelling, and show that an additional measurement at a shorter FIR wavelength can help constrain this. We present fitted relations between mass-weighted dust temperature and ratios of emission at two rest-frame wavelengths that can be applied widely to observational data.

\section*{Acknowledgements}
We thank Caleb Choban and Desika Narayanan for helpful comments on an earlier version of this Letter and Rachel Somerville for useful discussions. We also thank August Muench for help with the interactive figure. The Flatiron Institute is supported by the Simons Foundation. DAA was supported in part by NSF grants AST-2009687 and AST-2108944, and CXO grant TM2-23006X.

\bibliographystyle{aasjournal}
\bibliography{Edinburgh}

\end{document}